\RequirePackage{silence}
\documentclass[fleqn,usenatbib,useAMS]{mnras}
\usepackage{newtxtext,newtxmath}
\usepackage{graphicx}
\usepackage{amsmath}

\usepackage{amsfonts}
\usepackage{float}
\usepackage{bm}
\usepackage{ae,aecompl}
\usepackage{array}
\usepackage{soul}
\usepackage{mathtools}
\usepackage{multirow}
\usepackage[T1]{fontenc}
\usepackage[utf8]{inputenc}
\usepackage{booktabs}
\usepackage{graphicx}
\usepackage{subcaption}
\usepackage[normalem]{ulem}
\usepackage{anyfontsize}
\usepackage{xcolor}
\DeclareRobustCommand{\VAN}[3]{#2}
\let\VANthebibliography\thebibliography
\def\thebibliography{\DeclareRobustCommand{\VAN}[3]{##3}\VANthebibliography}

\DeclareRobustCommand{\appropto}{\mathrel{\vcenter{
		\offinterlineskip\halign{\hfil$##$\cr %The $##$ is not a mistake!
			\propto\cr\noalign{\kern2pt}\sim\cr\noalign{\kern-2pt}}}}}

\makeatletter
\DeclareRobustCommand*\bigcdot{\mathpalette\bigcdot@{2.4}}
\DeclareRobustCommand*\bigcdot@[2]{\mathbin{\vcenter{\hbox{\scalebox{#2}{$\m@th#1\bullet$}}}}}
\makeatother

\makeatletter
\DeclareRobustCommand*\diamond{\mathpalette\diamond@{1.6}}
\DeclareRobustCommand*\diamond@[2]{\mathbin{\vcenter{\hbox{\scalebox{#2}{$\m@th#1\blacklozenge$}}}}}
\makeatother

\title[The local void model for the Hubble and BAO tensions]{The local void model for the Hubble and BAO tensions} %Use [Short title]{Full title} if the two are inequivalent, former should be <=45 characters.

\author[I. Banik et al.]{Indranil Banik$^{1}$\thanks{E-mail: \href{mailto:indranil.banik@port.ac.uk}{indranil.banik@port.ac.uk} (Indranil Banik)}, Harry Desmond$^{1}$, Vasileios Kalaitzidis$^{2}$ and Sergij Mazurenko$^{3}$\\
$^{1}$Institute of Cosmology and Gravitation, University of Portsmouth, Dennis Sciama Building, Burnaby Road, Portsmouth PO1 3FX, UK\\
$^{2}$Scottish Universities Physics Alliance, University of Saint Andrews, North Haugh, Saint Andrews, Fife, KY16 9SS, UK\\
$^{3}$Universit\"{a}t Bonn, Regina-Pacis-Weg 3, D-53115 Bonn, Germany}

\date{Accepted XXX. Received YYY; in original form ZZZ}

\pubyear{\the\year{}}

\begin{document}
\label{firstpage}
\pagerange{\pageref{firstpage}--\pageref{lastpage}}
\maketitle

\begin{abstract} %No blank line below
The inconsistency between the locally inferred Hubble constant and the value inferred from the cosmic microwave background assuming the $\Lambda$CDM cosmological model has persisted, turning into an important problem. An emergent underlying trend is that this Hubble tension is driven by data confined to the very low-redshift Universe (typically $z < 0.15$). Most intermediate-redshift measurements remain mutually consistent with $H_0^\mathrm{CMB}$, the $\Lambda$CDM expectation anchored by the CMB. This Perspective examines if a large local void can explain the Hubble tension and its appearance only at low $z$. For an observer residing within a large underdensity, such as the Milky Way inside the claimed KBC void, gravitationally induced outflows and redshift can inflate the locally inferred recession scale $cz'$ despite having $H_0 = H_0^\mathrm{CMB}$. We summarise evidence suggestive of a local underdensity from multi-wavelength galaxy number counts, discuss the dynamical requirements implied by the amplitude of inferred bulk flows, and connect the solution to the emerging low-redshift BAO distance anomaly ($\alpha_{\mathrm{iso}} < 1$). Previously published semi-analytic void models anticipated the observed redshift dependence of BAO deviations and predict a rapid convergence to CMB-consistent expansion for $z \ga 0.2$, aligning with reconstructions of $H_0(z)$ from BAO plus uncalibrated Type~Ia supernovae. We conclude by looking to future tests, including improved mapping of the local density and velocity field, fits to galaxy distance catalogues at the field level, kinematic Sunyaev-Zel'dovich constraints on coherent outflows, fast radio bursts, and the long-term prospect of redshift drift measurements as a direct probe of time-varying non-cosmological redshift contributions.

\end{abstract}

%Consider citing this if it is published: https://arxiv.org/abs/2602.06293

% Select between one and six entries from the list of approved keywords. Do not make up new ones.

\section{The Hubble tension}
\label{Hubble_tension}

Cosmology is currently in a ``crisis'' known as the Hubble tension, a statistically significant mismatch between the rate at which redshift $z$ increases with distance $r$ in the local Universe, and the predicted rate of increase in the $\Lambda$ cold dark matter ($\Lambda$CDM) standard cosmological paradigm \citep*{Efstathiou_1990, Ostriker_Steinhardt_1995} with an early Universe calibration. If the Universe is taken to be homogeneous and isotropic, we must have that
\begin{eqnarray}
    cz' ~=~ \dot{a} = H_0 \, ,
    \label{Hubble_law}
\end{eqnarray}
where $c$ is the speed of light, $z' \equiv dz/dr$ in the local Universe, $a$ is the cosmic scale factor normalised to unity today, overdots denote time derivatives, 0 subscripts denote quantities at the present epoch, and $H \equiv \dot{a}/a$ is the Hubble parameter \citep*{Mazurenko_2025}. The assumption of homogeneity implies that the Hubble constant $H_0$ is a constant across space, but $H$ has almost certainly changed over time. It is possible to predict the form of $H(z)$ in the $\Lambda$CDM model once its parameters are calibrated in some way. By far the most precise calibration nowadays is provided by the anisotropies in the cosmic microwave background (CMB) radiation, which was emitted at $z = 1090$ when hydrogen in the early Universe underwent a phase transition from ionised plasma to neutral atoms as the expanding Universe adiabatically cooled. The CMB has characteristic temperature and polarisation fluctuations which encode a wealth of information about the early Universe. Just as we can learn the approximate size of a musical instrument purely by listening to its pitch, so also we can learn about the size of the early Universe from the power spectrum of the $\mathcal{O} (10^{-5})$ temperature fluctuations in the CMB. Nowadays, up to 10 peaks have been observed in the CMB power spectrum, providing detailed measurements of not only the `fundamental mode' but many of the `harmonics' \citep{Planck_2020, Tristram_2024, Calabrese_2025, Camphuis_2025}. In the case of the CMB, these arise due to fluctuations imprinted at early times on different length scales undergoing a different number of oscillations in the fixed time between the Big Bang and recombination. The first peak in the CMB power spectrum (on the largest angular scale) corresponds to an initially compressive density perturbation reaching maximum compression at the moment of recombination. Shorter wavelength modes reach this phase earlier, after which radiation pressure causes them to expand again. The second peak corresponds to an initial compression reaching maximum expansion at recombination, while the third peak corresponds to even smaller modes that reach maximum compression for a second time. Subsequent peaks arise for similar reasons. The most precise prediction of $H_0$ from analysing these results is $H_0^\mathrm{CMB} = 67.24 \pm 0.35$~km/s/Mpc \citep{Camphuis_2025}. This is based on combining CMB observations from the \emph{Planck} satellite \citep{Planck_2020} with those from the South Pole Telescope \citep[SPT;][]{Calabrese_2025, SPT_2025} and the Atacama Cosmology Telescope \citep[ACT;][]{Camphuis_2025}. Both SPT and ACT are on the ground, allowing them to reach smaller angular scales using larger instruments, thereby accessing further peaks in the power spectrum. This allows access to the Silk diffusion damping tail, where temperature fluctuations are smoothed by thermal conduction \citep{Silk_1968}. $\Lambda$CDM is able to fit this wealth of CMB data remarkably well, but only with particular values for the cosmological parameters, especially $H_0$.

This precise prediction for $H_0$ is in tension with measurements in the local Universe, which mostly find that $cz'$ is about 9\% higher \citep{Freedman_2001, Riess_2019, Valentino_2021_problem, Valentino_2025}. Recently, the $H_0$ Distance Network Collaboration \citep{H0DN_2025} reported a ``community consensus'' estimate of $cz' = 73.50 \pm 0.81$~km/s/Mpc, combining several popular methods and taking into account their covariances. This result is in $>7\sigma$ tension with $H_0^\mathrm{CMB}$. Results are consistent between a wide variety of techniques used to obtain the local $cz'$, not all of which rely on the traditional route of calibrating the absolute magnitude of Type~Ia supernovae (SNe~Ia) with the Leavitt law \citep{Leavitt_1912} relating periods and luminosities of Cepheid variable stars \citep{Scolnic_2023}. For instance, it is possible to obtain 1.8\% precision using Cepheids and geometric anchors alone \citep{Stiskalek_2026_Cepheids}, but even this result is $3.3\sigma$ above $H_0^\mathrm{CMB}$. There is also good agreement between distances to nearby galaxies obtained with the \emph{Hubble Space Telescope} \citep[\emph{HST};][]{Riess_2022_comprehensive} and the subset that were recently followed up using the larger and longer wavelength \emph{James Webb Space Telescope} \citep[\emph{JWST};][]{Riess_2024_consistency, Freedman_2025, Li_2025}. Despite finding distances consistent with earlier studies, a low $H_0$ was reported using a limited number of SNe~Ia observed by \emph{JWST} \citep{Freedman_2025}. This is due to sample size differences -- restricting \emph{HST} data to the same SNe~Ia would also give a low $H_0$ \citep{Riess_2024_consistency}. These results make it unlikely that modern distance estimates underpinning the Hubble tension are underestimated due to the historically important issue of crowding of Cepheid variables in telescopes with limited resolution \citep{Riess_2024_crowding}. Moreover, the Hubble tension is apparent without Cepheids or SNe~Ia, for instance using Type~II supernovae (SNe~II) without external calibration \citep{Vogl_2025}, or using a distance ladder approach based on the tip of the red giant branch (TRGB) and surface brightness fluctuations \citep[SBF;][]{Jensen_2025}. A $2\sigma$ tension with $H_0^\mathrm{CMB}$ is apparent from purely geometric distances to megamaser host galaxies, which give $cz' = 73.9 \pm 3.0$~km/s/Mpc \citep{Pesce_2020}.

\begin{figure*}
    \centering
    \includegraphics[width=\textwidth]{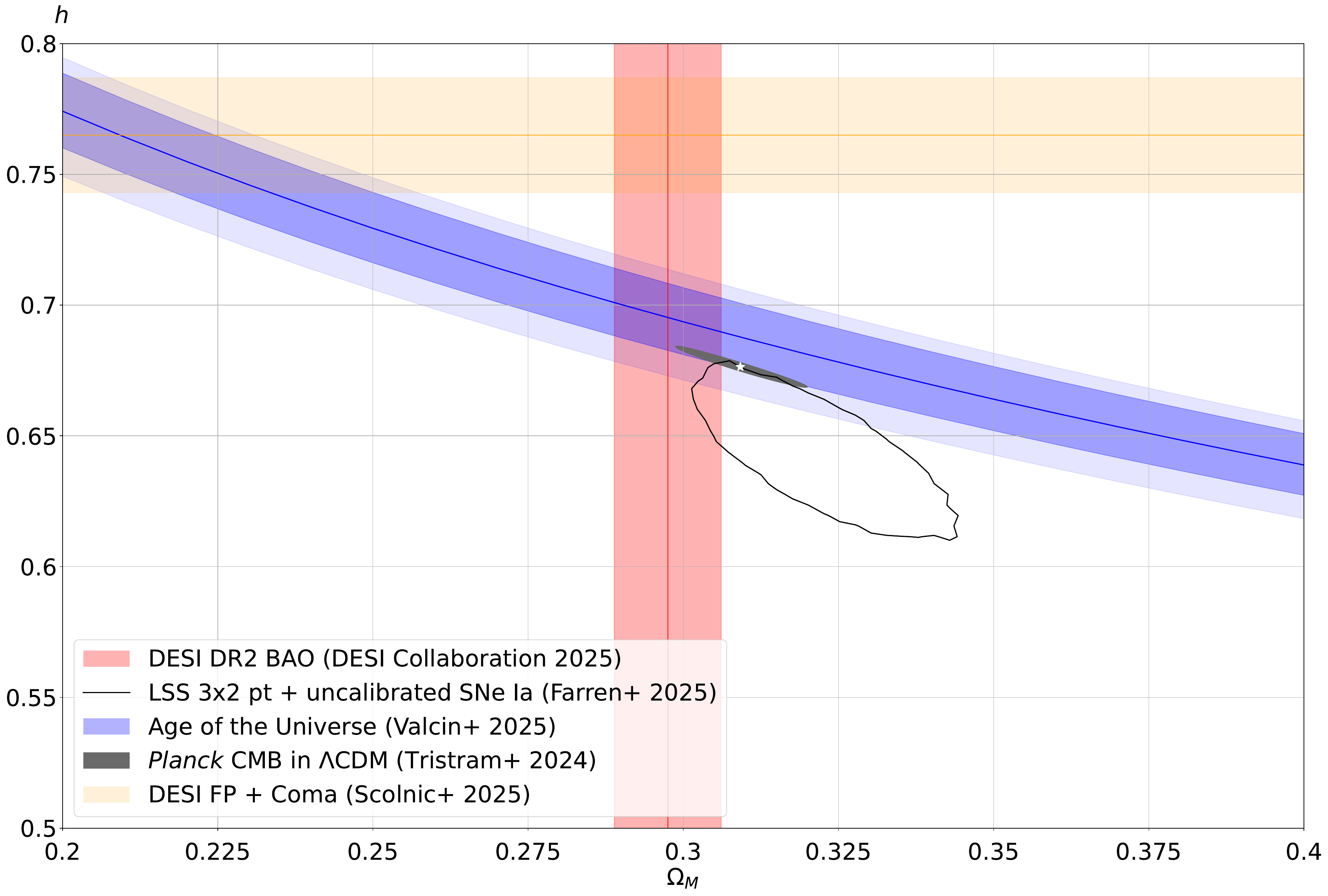}
    \caption{Constraints on $h \equiv H_0$ in units of 100 km/s/Mpc and $\Omega_\mathrm{M}$, the fraction of the cosmic critical density presently in the matter component. The solid lines show the most likely values, while the shaded bands show $1\sigma$ uncertainties. The CMB constraint assuming $\Lambda$CDM is shown as the thin grey ellipse with central white star \citep{Tristram_2024}. The horizontal orange band shows a typical local $cz'$ measurement \citep{Said_2025, Scolnic_2025}. The tension between these two is the Hubble tension. The vertical red band obtains $\Omega_\mathrm{M}$ using uncalibrated BAO data from DESI~DR2 \citep{DESI_2025}. The sloped blue band uses the age of the Universe from old Galactic globular clusters \citep{Valcin_2025}. The dark blue region assumes the oldest one formed $t_\mathrm{f} = 0.2$~Gyr after the Big Bang, while the light blue band allows $t_\mathrm{f}$ to be $2\times$ smaller or larger. The open black contour uses measurements of LSS and uncalibrated SNe~Ia \citep{Farren_2025}. Notice how the local $H_0$ is the only constraint not consistent with the others. Reproduced from figure~2 of \citet{Banik_2025_cosmology}.}
    \label{OmegaM_h}
\end{figure*}

In the $\Lambda$CDM theory, we can obtain $H_0$ from other kinds of probes at redshifts intermediate between the CMB and the $z < 0.15$ range used in local measurements. Some of the most important such results are summarised in Figure~\ref{OmegaM_h}, which shows constraints on $H_0$ and $\Omega_\mathrm{M}$, the fraction of the cosmic critical density presently in the matter component (including neutrinos). The Hubble tension is apparent as the mismatch between the horizontal orange band showing one estimate of the local $cz'$ \citep{Scolnic_2025} and the thin grey ellipse with central white star showing the CMB constraint \citep{Tristram_2024}. The blue band uses the age of the Universe ($A_\mathrm{U}$), which is found by adding $t_\mathrm{f} = 0.2$~Gyr to the age limit inferred for the Galactic globular cluster population \citep{Valcin_2025}. The dark blue band fixes the formation time $t_\mathrm{f}$, while the light blue band allows $t_\mathrm{f}$ to be $2\times$ smaller or larger, with higher $t_\mathrm{f}$ moving the constraint down to lower $H_0$. The vertical red band uses uncalibrated baryon acoustic oscillation (BAO) data from the Dark Energy Spectroscopic Instrument Data Release 2 \citep[DESI~DR2;][]{DESI_2025}. Since the comoving length $r_\mathrm{d}$ of the BAO ruler is left unconstrained, we cannot obtain $H_0$, but we can precisely constrain $\Omega_\mathrm{M}$. More generally, uncalibrated cosmic standards (UCS) provide constraints on $a(t/A_\mathrm{U})$, the shape of the cosmic expansion history \citep*{Lin_2021_UCS}. Uncalibrated SNe~Ia are combined with large-scale structure (LSS) for the open black contour, which uses the galaxy-galaxy, galaxy-lensing, and lensing-lensing cross-correlation power spectrum, with lensing here referring to weak lensing \citep{Farren_2025}. Similar results were reported using DESI \citep{Zaborowski_2025}. A key goal of such studies is to find the turnover scale in the matter power spectrum corresponding to the comoving horizon at matter-radiation equality, when the universe was about $3\times$ smaller and $7\times$ younger than at recombination \citep{Banik_2025_cosmology}. This scale is imprinted on LSS because modes entering the horizon at earlier epochs had to wait until the era of matter domination before density perturbations could start growing \citep{Meszaros_1974, Eisenstein_1998}.

Figure~\ref{OmegaM_h} shows that all the datasets are consistent in $\Lambda$CDM -- except for the local $cz'$. This is the only constraint based solely on data at $z < 0.15$, a region to which the other techniques have very little sensitivity as they are based on the more distant Universe. The excellent agreement between the CMB constraint and other intermediate redshift probes indicates that the Hubble tension is best thought of as a mismatch between the local $cz'$ and other $\Lambda$CDM-based estimates of $H_0$ from outside the local Universe \citep{Lin_2021_UCS, Perivolaropoulos_2024, Pantos_2026}.

The good agreement between local $cz'$ measurements by different teams using different techniques and instruments makes it unlikely that distances in the local Universe are consistently underestimated by 9\%. At the same time, $\Lambda$CDM requires $H_0 \approx H_0^\mathrm{CMB}$ in order to explain several aspects of the $z \ga 0.15$ Universe, even without considering the information encoded in CMB anisotropies \citep{Banik_2025_cosmology}. $\Lambda$CDM therefore makes a very clear prediction for $H_0$. The apparent failure of $\Lambda$CDM to predict this most basic of cosmological parameters has motivated a wide variety of suggestions in the literature \citep{Valentino_2021_solutions, Valentino_2025}. The vast majority of these proposals retain the validity of Equation~\ref{Hubble_law}, instead seeking to raise $H_0$ while maintaining consistency with the CMB. However, it is possible that $H_0 = H_0^\mathrm{CMB}$ if $cz' > H_0$ in the local Universe. This requires part of the redshift to arise from sources other than homogeneous cosmic expansion.

\section{The local void solution}
\label{Local_void_solution}

Non-cosmological contributions to the observed redshift are all around us. The most obvious source is the peculiar velocity of a galaxy arising from nearby galaxies and LSS \citep{Watkins_2025}. For instance, M31 is approaching our Galaxy \citep{McConnachie_2012}. This implies their mutual gravity turned around their initial expansion within a Hubble time, constraining their total mass via the classic `timing argument' \citep{Kahn_Woltjer_1959}. Besides this indirect contribution of gravity to the redshift via induced peculiar velocities and the Doppler effect, there is also a direct contribution through gravitational redshift (GR). The latter has been detected around the Galactic centre black hole \citep{Gravity_2018} and active galactic nuclei \citep{Padilla_2024}. These effects are especially important when inferring $cz'$ from nearby sources, since their small cosmological redshift increases the relative importance of any non-cosmological contribution \citep{Carneiro_2022}.

To explain the Hubble tension through an inflated local value of $cz'$, the observed redshift must increase with distance more rapidly than expected from cosmic expansion alone. This requires additional contributions to the redshift -- either from coherent outward peculiar velocities relative to the Hubble flow, or from GR due to our location on a potential hill. Both effects arise naturally in an underdense environment: a local void produces outflowing peculiar velocities and places the observer at higher gravitational potential relative to distant sources. Although peculiar velocities would not continue increasing indefinitely with distance if the Universe is homogeneous on large scales, an increasing trend over a limited range of distances is entirely plausible. If the void properties and our vantage point are carefully chosen, this may be sufficient to explain the high local $cz'$. An obvious expectation is that the Hubble tension should not persist to high redshift. This is consistent with the results shown in Figure~\ref{OmegaM_h}.

A local void was proposed long before the Hubble tension based on optical galaxy number counts \citep{Maddox_1990, Shanks_1990}. A clearer picture emerges from near-infrared observations covering 90\% of the sky, which suggest that we live inside the Keenan-Barger-Cowie (KBC) void \citep*{Keenan_2013}. This structure is suggested by evidence across the electromagnetic spectrum, from X-rays \citep{Bohringer_2015, Bohringer_2020} to radio wavelengths \citep*{Rubart_2013, Rubart_2014} and in the infrared \citep{Huang_1997, Busswell_2004, Frith_2003, Frith_2005, Frith_2006, Keenan_2013, Whitbourn_2014, Whitbourn_2016, Wong_2022}. The assumed value of $cz'$ does not affect these results because the underlying observable is the galaxy luminosity density in redshift space. The increasing comoving luminosity density at larger $z$ is difficult to understand through selection effects, which would instead be expected to favour the detection of nearby galaxies. To explain the observations in a homogeneous universe, extreme evolution of the galaxy population would be required at $z \la 0.1$ followed by much milder evolution out to $z = 1$ \citep{Wong_2022}. Taking the near-infrared observations at face value, comparison with the Millennium XXL simulation \citep{Angulo_2012} shows that the KBC void is $6\sigma$ discrepant with $\Lambda$CDM due to its large size and depth \citep*{Haslbauer_2020}. Consequently, any solution to the Hubble tension not appreciably altering the growth of structure on scales $\ga 100$~Mpc would likely struggle to explain the existence of the KBC void.

Several authors have argued that the Hubble tension could be solved by gravitationally driven outflows from the KBC void \citep{Keenan_2016, Shanks_2019a, Shanks_2019b, Ding_2020, Martin_2023}. To estimate its impact on $cz'$, we first note that the KBC void has a claimed underdensity in redshift space of $\delta_{\mathrm{obs}} = 46 \pm 6\%$ \citep[see figure~11 of][]{Keenan_2013}. The actual underdensity is probably around half as much due to redshift space distortion \citep{Kaiser_1987}, the reduction in distance to any fixed redshift due to outflow from a local void inflating $cz'$. This reduces the volume enclosed within any fixed $z$, reducing the number counts and making the underdensity appear deeper in redshift space than it actually is. Accounting for this effect and noting that the required reduction in comoving density from the nearly homogeneous early Universe must be due to an increase in the KBC void's comoving volume, \citet{Haslbauer_2020} argued in their equation~5 that
\begin{eqnarray}
    \frac{cz'}{H_0} ~=~ \left( 1 - \delta{_\mathrm{obs}} \right)^{-1/6}.
    \label{H0_estimate_void}
\end{eqnarray}
This simple estimate implies that $cz'$ should exceed $H_0^\mathrm{CMB}$ by $11 \pm 2\%$. The excess will depend on details of the void's formation history, since the continuity equation implies that its impact on $cz'$ is directly related to the present rate at which the void is becoming deeper. The growth history of the KBC void must differ from how voids typically grow in $\Lambda$CDM, which cannot form such a large and deep void given the initial conditions set by the CMB \citep{Haslbauer_2020}.

The KBC void has a radius of about 300~Mpc. Fluctuations on this comoving scale are extremely well constrained from observations of the CMB, which shows excellent agreement with $\Lambda$CDM expectations \citep{Tristram_2024}. However, this only constrains the fluctuations early in cosmic history. The local void solution assumes that density perturbations on physical scales $\ga 100$~Mpc grow faster than in $\Lambda$CDM. We suggest that this is due to a modification to gravity on the corresponding length scales. Such a modification would not affect the early universe as its smaller age reduced the cosmic horizon, the maximum distance that light and gravity could have travelled \citep{Haslbauer_2020}.

Solving the Hubble tension through a local void requires peculiar velocities that are far larger than expected in $\Lambda$CDM. This is in apparent agreement with the bulk flow curve out to $z = 0.083$ (about 350~Mpc) inferred by \citet{Watkins_2023} based on applying the minimum variance estimator \citep*{Peery_2018} to the Cosmicflows-4 (CF4) galaxy catalogue \citep{Tully_2023}. The bulk flow measures the average vector velocity of tracers within a spherical region centred on the Sun. A local void explanation typically requires coherent peculiar velocities that are larger than the median $\Lambda$CDM expectation on comparable scales. Bulk flow measurements are suggestive in this direction, but their interpretation can be sensitive to distance error modelling, selection effects, and how correlated errors are handled. \citet{Watkins_2023} report that the bulk flow amplitude on their largest probed scale of $250/h$~Mpc is roughly quadruple the $\Lambda$CDM expectation under their modelling assumptions; the formal significance can be reduced if distance uncertainties and other catalogue systematics are larger than assumed. Their results were later supported by \citet{Whitford_2023}, who concluded that the ``bulk flow amplitude is in excellent agreement with the [earlier] measurement''. However, they also found that the uncertainties may have been underestimated, which if true would reduce the tension with $\Lambda$CDM \citep{Whitford_2023}. Besides the high amplitude of the reported bulk flows, another unusual aspect is that they increase with distance, a behaviour which is not trivially explainable within the $\Lambda$CDM framework -- but arises naturally for an observer slightly offset from the centre of a roughly spherically symmetric KBC-like void \citep{Mazurenko_2024}.

A bulk flow is essentially a dipole in $z$ at fixed $r$, so it can also cause $cz'$ to become anisotropic. Some studies do indeed report such anisotropy and find a consistent preferred direction \citep{Kalbouneh_2023, Hu_2024_RF, Hu_2024_Pade, Kalbouneh_2025, Salzano_2025}. However, a recent detailed analysis found no evidence for anisotropy in the local $cz'$ \citep{Stiskalek_2026_anisotropy}. This issue is currently being investigated using SNe~Ia (Mazurenko et al., submitted). A direct confirmation of the large peculiar velocities expected in the local void solution therefore remains elusive (Section~\ref{kSZ_effect}).

\section{Relation to the BAO anomaly}
\label{BAO_anomaly_relation}

The very prominent first acoustic peak in the CMB power spectrum implies that matter is preferentially clustered on a characteristic scale at recombination. In $\Lambda$CDM, this characteristic scale is the sound horizon of size $r_\mathrm{d} \approx 147$ comoving Mpc (cMpc). An excess clustering amplitude on this scale is therefore expected in LSS \citep{Eisenstein_1998}. Indeed, the galaxy power spectrum does have a characteristic `bump' known as the BAO feature, which was first detected in 2005 \citep{Cole_2005, Eisenstein_2005}. The comoving size of this feature is $r_\mathrm{d}$, which was fixed since $z_\mathrm{d} \approx 1060$ once photons and baryons decoupled. This BAO standard ruler can be oriented parallel or orthogonal to the line of sight (LOS), providing a redshift depth or angular scale, respectively. BAO measurements now span a wide range of redshifts, providing a crucial constraint on the expansion history at intermediate epochs \citep{Chen_2024_BAO}.

By combining angular and LOS data, we can estimate the volume of the `BAO bubble' and thereby obtain an isotropically averaged BAO distance scale $D_\mathrm{V}$, or more precisely the ratio $D_\mathrm{V}/r_\mathrm{d}$. In a homogeneous cosmology,
\begin{eqnarray}
    D_{\mathrm{V}} ~=~ \left( z D_{\mathrm{c}}^2 D_{\mathrm{H}} \right)^{1/3} \, ,
    \label{D_V}
\end{eqnarray}
where $D_H \equiv c/H$ at any redshift $z$, the comoving distance to which is $D_{\mathrm{c}} = c \int_0^z d\tilde{z}/H(\tilde{z})$. Instead of reporting $D_V/r_\mathrm{d}$ directly, it is common to report its ratio $\alpha_\mathrm{iso}$ with the expected value in some fiducial cosmology.
\begin{eqnarray}
    \alpha_{\mathrm{iso}} ~\equiv~ \frac{D_{\mathrm{V}}}{r_{\mathrm{d}}} \div \left. \frac{D_{\mathrm{V}}}{r_{\mathrm{d}}} \right|_{\mathrm{fid}} ~=~ \left( \alpha_\perp^2 \alpha_\parallel \right)^{1/3} \, ,
    \label{alpha_iso}
\end{eqnarray}
where the `iso' subscript indicates an isotropically averaged value and `fid' subscripts indicate values in the fiducial cosmology. This is usually taken to be the \emph{Planck} cosmology \citep{Planck_2020}, a convention we follow here. It is possible to use only the angular BAO scale to obtain $\alpha_\perp$ or only its redshift depth to obtain $\alpha_\parallel$, both of which are defined analogously to $\alpha_{\mathrm{iso}}$. Combining these into $\alpha_{\mathrm{iso}}$ improves the accuracy and increases the number of available BAO measurements, since many studies only report $\alpha_{\mathrm{iso}}$ \citep{Banik_2025_BAO}. Those authors consider all these types of $\alpha$ and $\alpha_{_{\mathrm{AP}}} \equiv \alpha_\perp/\alpha_\parallel$ \citep{Alcock_1979}, which can be particularly useful for testing void models in the future \citep{February_2013, Clarkson_2012}. Besides $\alpha_{\mathrm{iso}}$ which we focus on here, the most useful quantity is $\alpha_\perp$ because the BAO angular scale is usually easier to measure than its redshift depth.

The low-redshift nature of the Hubble tension (Figure~\ref{OmegaM_h}) implies a low-redshift distortion to the distance-redshift relation predicted by $\Lambda$CDM. This must inevitably show up as some kind of anomaly in the BAO data. If the predicted $r_\mathrm{d}$ is correct but the local $cz'$ exceeds the $\Lambda$CDM prediction, then the distance to any fixed redshift will be smaller than in $\Lambda$CDM, implying that $\alpha_{\mathrm{iso}} < 1$. BAO measurements over the last 20 years show just such a BAO anomaly \citep{Beutler_2011, Carter_2018, Carvalho_2021, Tully_2023_BAO, DES_2024_BAO, DESI_2025, DES_2026}, as reviewed in \citet{Banik_2025_BAO}. Their table~2 shows that the discrepancy with $\Lambda$CDM has now reached about $3.3\sigma$ based on $D_V$, but this misses a few studies that only report the angular BAO scale. Including them raises the tension to $3.8\sigma$. These are approximate levels of tension based on combining all the available BAO measurements. Although some studies were excluded because their data were later reused in a different study, covariances between BAO measurements are not rigorously included and are very difficult to calculate. However, a $2.9\sigma$ tension is apparent from DESI~DR2 alone (see their section~3.2). Other studies report that the BAO anomaly has a similar level of significance using more rigorous methods \citep{Camphuis_2025, Najera_2026}.

\begin{figure*}
    \centering
    \includegraphics[width=\textwidth]{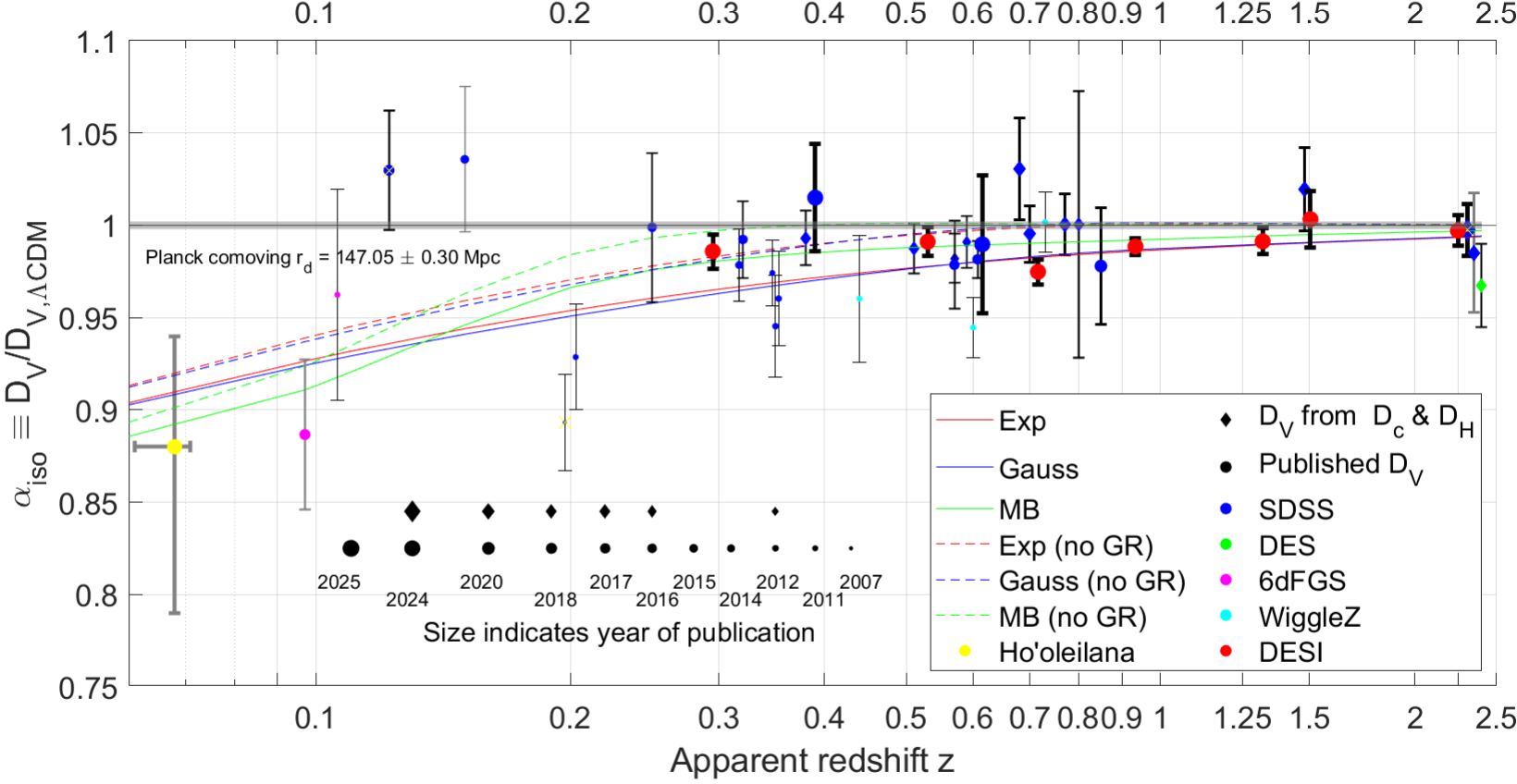}
    \caption{BAO measurements of $\alpha_{\mathrm{iso}}$ (Equation~\ref{alpha_iso}) over the last 20~years, scaled to the prediction in the \emph{Planck} cosmology. The colour of each point indicates the survey used, while the marker size indicates the year of publication, emphasizing more recent measurements. The $\Lambda$CDM prediction is by definition unity at all $z$, with the approximate uncertainty shown by the shaded grey band. Predictions of the void models are shown with solid coloured lines for different initial underdensity profiles, as indicated in the legend. For illustrative purposes, the dashed lines show the corresponding model predictions without GR (Equation~\ref{z_contributions}), highlighting its importance. To avoid double counting studies whose observations were later reused in a different study shown here, the points with grey error bars were excluded when quantitatively comparing model predictions to observations. Reproduced from figure~4 of \citet{Banik_2025_BAO}.}
    \label{D_V_graph}
\end{figure*}

Figure~\ref{D_V_graph} shows $\alpha_{\mathrm{iso}}$ measurements over the last 20 years. The BAO anomaly is clearly evident here as a growing departure from $\alpha_\mathrm{iso} = 1$ at $z \la 1$. Importantly for the Hubble tension, the anomaly is towards smaller distances to a given redshift than in the \emph{Planck} cosmology ($\alpha_{\mathrm{iso}} < 1$). This is what we would expect if the local $cz'$ is anomalously large, as indicated by a substantial body of evidence \citep{H0DN_2025, Valentino_2025}. The BAO anomaly is therefore unsurprising. Indeed, the BAO data can be predicted remarkably well if we constrain the $\Lambda$CDM cosmological parameters using both the local $cz'$ and the CMB, even though $\Lambda$CDM cannot simultaneously fit both datasets \citep{Najera_2026}. If the local $cz'$ measurements suffer from significant systematics, constraining $\Lambda$CDM parameters using these measurements in conjunction with the CMB might be expected to yield incorrect BAO predictions, given BAO data has very different systematics.

Figure~\ref{D_V_graph} also highlights the predictions of the local void model, with solid lines showing the results for initially exponential, Gaussian, and Maxwell-Boltzmann (MB) underdensity profiles \citep{Haslbauer_2020}. Those authors set up a semi-analytic model of a spherical void by starting with a small initial underdensity and evolving a grid of test particles under the assumption that the enclosed mass remains constant. Peculiar accelerations are driven by the mass deficit, i.e., how much the enclosed mass falls below the mass in the same volume if the universe were homogeneous. Since $\Lambda$CDM cannot account for the reported properties of the KBC void, those authors boosted the Newtonian gravity using an extra factor inspired by Milgromian dynamics \citep[MOND;][]{Milgrom_1983}. The test particles were evolved from $z = 9 \to 0$ assuming a background \emph{Planck} cosmology. The void parameters and the external field strength (which regulates the enhancement to gravity) were then constrained for each profile using a variety of observations, mainly the local $cz'$ and galaxy number counts. Importantly, BAO data were not considered in the fits.

To extract BAO predictions, it is necessary to consider our past lightcone. Since BAO data comes from quite high redshift, a valid simplifying approximation is that our location is at the void centre. The distance-redshift relation perceived by an observer here is defined by the locus of points where their past lightcone intersects the trajectory of each test particle. It will appear redshifted to some extent by cosmic expansion over the light travel time, but a local void introduces additional contributions to the redshift. In particular, equation~52 of \citet{Haslbauer_2020} shows that the observed redshift $z$ satisfies:
\begin{eqnarray}
    1 + z ~=~ \frac{1}{a \left( t \right)} \overbrace{\sqrt{\frac{c + v_{\mathrm{int}}}{c - v_{\mathrm{int}}}}}^{\text{Doppler}}  \overbrace{\exp \left( \frac{1}{c^2} \int g_{\mathrm{void}} \, dr \right)}^{\text{GR}} \, ,
    \label{z_contributions}
\end{eqnarray}
where $a$ was the cosmic scale factor at the emission time $t$ of an observed photon, $g_{\mathrm{void}}$ is the outward gravitational field induced by the void due to it having less density than the cosmic mean, and $v_{\mathrm{int}}$ is the outflow velocity in the reference frame of the void, which may be moving as a whole. We neglect any such systemic velocity because it would cancel out if observing sources at the same $z$ over a wide range of sky directions, allowing us to deal only with $v_{\mathrm{int}}$.

The distance-redshift relation is distorted by the non-cosmological redshift induced by outflows from a local void and its outward gravitational push on incoming photons. The predicted run of $\alpha_{\mathrm{iso}}$ is shown in Figure~\ref{D_V_graph}, which uses the same best-fitting parameters for each void profile as reported by \citet{Haslbauer_2020}. Model predictions are shown with solid lines, while dashed lines in the same colour show the impact of excluding GR (Equation~\ref{z_contributions}). At the high redshifts covered by BAO measurements, the GR contribution is much more important than the Doppler effect from outward peculiar velocities, which we expect would be small at such large distances.

Since the local void models were constrained using the local $cz'$, the void models reach $\alpha_{\mathrm{iso}} \approx 0.91$ as $z \to 0$. At larger redshifts, the decaying impact of a local void implies $\alpha_{\mathrm{iso}} \to 1$. Figure~\ref{D_V_graph} shows that the rate at which this transition occurs agrees very well with the observations for all three void profiles. Since neither the fiducial cosmology nor the void models were constrained using BAO data, these can test their \emph{a priori} predictions. The void models are substantially preferred \citep{Banik_2025_BAO}. Considering all 42 available $D_V/r_\mathrm{d}$ measurements (excluding duplicates where appropriate), those authors found that the total $\chi^2$ decreases from 76 down to a much more reasonable $47-51$, depending on the void density profile (see their table~2). Including an additional 7 measurements of $D_c/r_\mathrm{d}$ from studies that only report this quantity and are therefore omitted in the above calculation, the total $\chi^2$ rises to 93 for the homogeneous \emph{Planck} cosmology. With only 49 data points, this represents a $3.8\sigma$ tension. The void models reduce $\chi^2$ to just $55-57$, implying a tension of between $1.1\sigma$ and $1.3\sigma$. Even when restricting the analysis to the 13 data points from DESI~DR2, the void models still reduce the tension from $2.9\sigma$ to $<2\sigma$, giving them a larger model probability in a Bayesian sense by a factor of about 200 (see their section~3.2).

These results demonstrate that the local void solution with a background \emph{Planck} cosmology predicted the latest BAO data using model parameters published several years earlier without considering BAO data \citep{Haslbauer_2020, Banik_2025_BAO}. The equivalent predictions made by the homogeneous \emph{Planck} cosmology are in significant tension with available BAO measurements over the last 20~years, even when these are limited to DESI~DR2. This success of the local void solution does not uniquely imply its validity, since a purely background solution to the Hubble tension that modifies the $\Lambda$CDM expansion history only at low $z$ might be expected to yield similar BAO predictions \citep{Najera_2026}. Those authors considered several background cosmologies that differ from $\Lambda$CDM at low $z$, but none of the models provide a particularly good joint solution to the Hubble and BAO tensions. Even so, a homogeneous background solution can in principle yield any distance--redshift relation -- including that predicted by a void model. Distinguishing background solutions from the local void scenario is thus challenging but potentially feasible, as we discuss in Section~\ref{Future_tests}.

\section{Redshift dependence of the Hubble tension in the apparent expansion history}
\label{Apparent_expansion_history}

\begin{figure*}
    \centering
    \includegraphics[width=\textwidth]{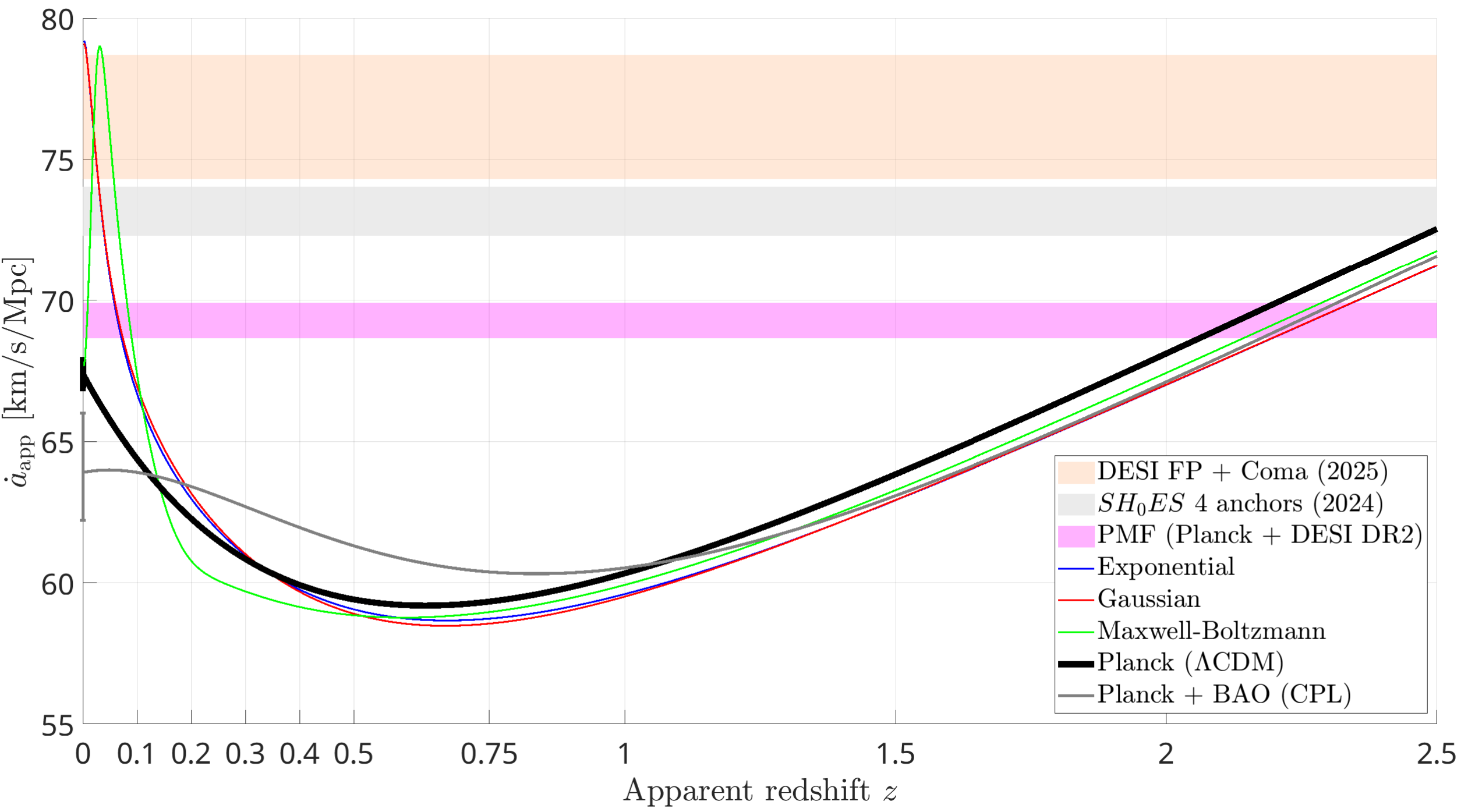}
    \caption{The cosmic expansion history in different models. The black curve shows the \emph{Planck} cosmology, while the grey curve shows the CPL model fit to CMB + BAO \citep{Mirpoorian_2025}. Uncertainties on the $H_0$ predictions of these models are indicated at the left. The lowest horizontal band in magenta shows the predicted $H_0$ in their PMF-inspired model where recombination occurs slightly earlier than in $\Lambda$CDM. The higher horizontal bands show two typical local $cz'$ measurements \citep{Breuval_2024, Scolnic_2025}. The coloured solid curves show predictions in the local void models, as indicated in the legend. The apparent $a$ at any lookback time is obtained from the measured redshift assuming homogeneity (Equation~\ref{a_app}). The true background $\dot{a}(z)$ in the void models is the solid black curve.}
    \label{adot_graph}
\end{figure*}

The BAO anomaly could in principle indicate that the cosmic expansion history $\dot{a}(z)$ differs from $\Lambda$CDM expectations. This may be due to time evolution of the dark energy density. Its equation of state $w$ varies linearly with $a$ in the Chevallier-Polarski-Linder (CPL) parametrisation \citep{Chevallier_2001, Linder_2003}. This CPL model has two extra free parameters compared to $\Lambda$CDM. The thin grey line on Figure~\ref{adot_graph} shows the CPL model that best matches BAO + CMB data. The error bar at $z = 0$ illustrates the uncertainty on the predicted $H_0$, which is not used to fix the parameters \citep*{DESI_2025, Mirpoorian_2025}. The CPL model predicts $H_0 = 64 \pm 2$~km/s/Mpc. The solid black line showing the \emph{Planck} cosmology predicts a higher $H_0$ with a far smaller uncertainty. Even if local $cz'$ measurements are not taken into account, the Bayesian Evidence still disfavours the CPL model given its greater complexity than $\Lambda$CDM \citep*{Ong_2025}.

Care must be taken when obtaining the apparent $\dot{a}(z)$ predicted by the local void models discussed previously. Observers typically report $\dot{a}$ or $H$ at some redshift assuming homogeneity, which is no longer valid. To show the apparent $\dot{a}(z)$ in the void models, we employ the same test particle integration scheme and lightcone analysis as discussed in Section~\ref{BAO_anomaly_relation} \citep{Haslbauer_2020}. Although redshifts in the void model arise from several sources, observers typically consider only the homogeneous contribution (the $1/a$ term in Equation~\ref{z_contributions}). Under this assumption, they would infer an apparent scale factor of
\begin{eqnarray}
    a_{\mathrm{app}} ~\equiv~ \frac{1}{1 + z} \, .
    \label{a_app}
\end{eqnarray}
By evaluating $a_{\mathrm{app}}$ for a grid of test particles along our past lightcone and recording the corresponding lookback times, it is possible to obtain $\dot{a}_{\mathrm{app}} (z)$.

The results are shown as thin coloured lines on Figure~\ref{adot_graph}. At $z = 0$, these reach a high $\dot{a}_{\mathrm{app}}$ consistent with the locally measured $cz'$. This is to be expected given the models were designed to solve the Hubble tension. Interestingly, the predictions are almost identical to those of the \emph{Planck} cosmology for $z > 0.2$. This emphasizes the local nature of the void solution. The predicted $\dot{a}_{\mathrm{app}} (z)$ is most directly related to cosmic chronometer (CC) measurements of $\dot{z}$ along our past lightcone using relative ages of passively evolving galaxies at different redshifts \citep{Jimenez_2002, Moresco_2018, Moresco_2020, Moresco_2024}. CCs are not yet sensitive enough to probe the $z \la 0.2$ regime, where a very recent measurement at $z = 0.12$ using several thousand DESI galaxies still has a 6\% uncertainty \citep{Wang_2026}. Future improvements will allow CCs to distinguish the various predicted $\dot{a}_{\mathrm{app}} (z)$ curves at $z \la 0.2$, especially since CC uncertainties appear to be underestimated \citep*{Kvint_2025}. However, if $c$ has remained constant, measuring $\dot{z}$ does not provide any extra information beyond that provided by $cz'$, for which there are already many published values with percent-level uncertainty and good mutual consistency (Section~\ref{Hubble_tension}).

At the moment, the most helpful aspect of CCs is their ability to check if the $z \ga 0.2$ Universe does indeed follow the \emph{Planck} cosmology, as expected in the local void solution. The results are in line with this prediction \citep{Cogato_2024, Niu_2026}. A careful combination of CCs with uncalibrated BAO and SNe~Ia datasets gave $H_0 = 68.4^{+1.0}_{-0.8}$~km/s/Mpc \citep{Guo_2025}. The close agreement with $H_0^\mathrm{CMB}$ is to be expected given the considered data is mostly at $z > 0.2$. These results argue against a scenario in which the Hubble tension is solved through new physics only prior to recombination, which in principle could allow a good fit to the CMB at higher $H_0$. Regardless of how such an early time solution operates, the lack of new physics in the late universe implies that $\dot{a}(z)$ should follow the same shape as the \emph{Planck} cosmology -- but scaled up by 9\% at all redshifts. CC datasets appear to contradict this prediction. A related problem is that early time solutions would also reduce $A_\mathrm{U}$ from the usual 13.8~Gyr down to $\approx 13.8/1.09 = 12.7$~Gyr. This is inconsistent with the ages of the oldest Galactic stars and globular clusters \citep{Valcin_2021, Xiang_2022, Cimatti_2023, Souza_2024, Lundkvist_2025, Shariat_2025, Tomasetti_2025, Valcin_2025, Xiang_2025}.

Early time solutions to the Hubble tension also face other issues \citep{Vagnozzi_2023}, including the lack of any anomalies in the CMB power spectrum that might be associated with the hypothesised new physics prior to recombination. This issue prevents a solution to the Hubble tension through a flexible modification to the ionisation history of the universe around the time of recombination inspired by primordial magnetic fields \citep[PMF;][]{Mirpoorian_2025}. This model increases the redshift of recombination from the standard 1090 up to about 1110. The predicted $H_0$ is shown on Figure~\ref{adot_graph} using the horizontal magenta band. This is still well below the local $cz'$, as shown by the higher horizontal bands.

\begin{figure*}
    \centering
    \includegraphics[width=\textwidth]{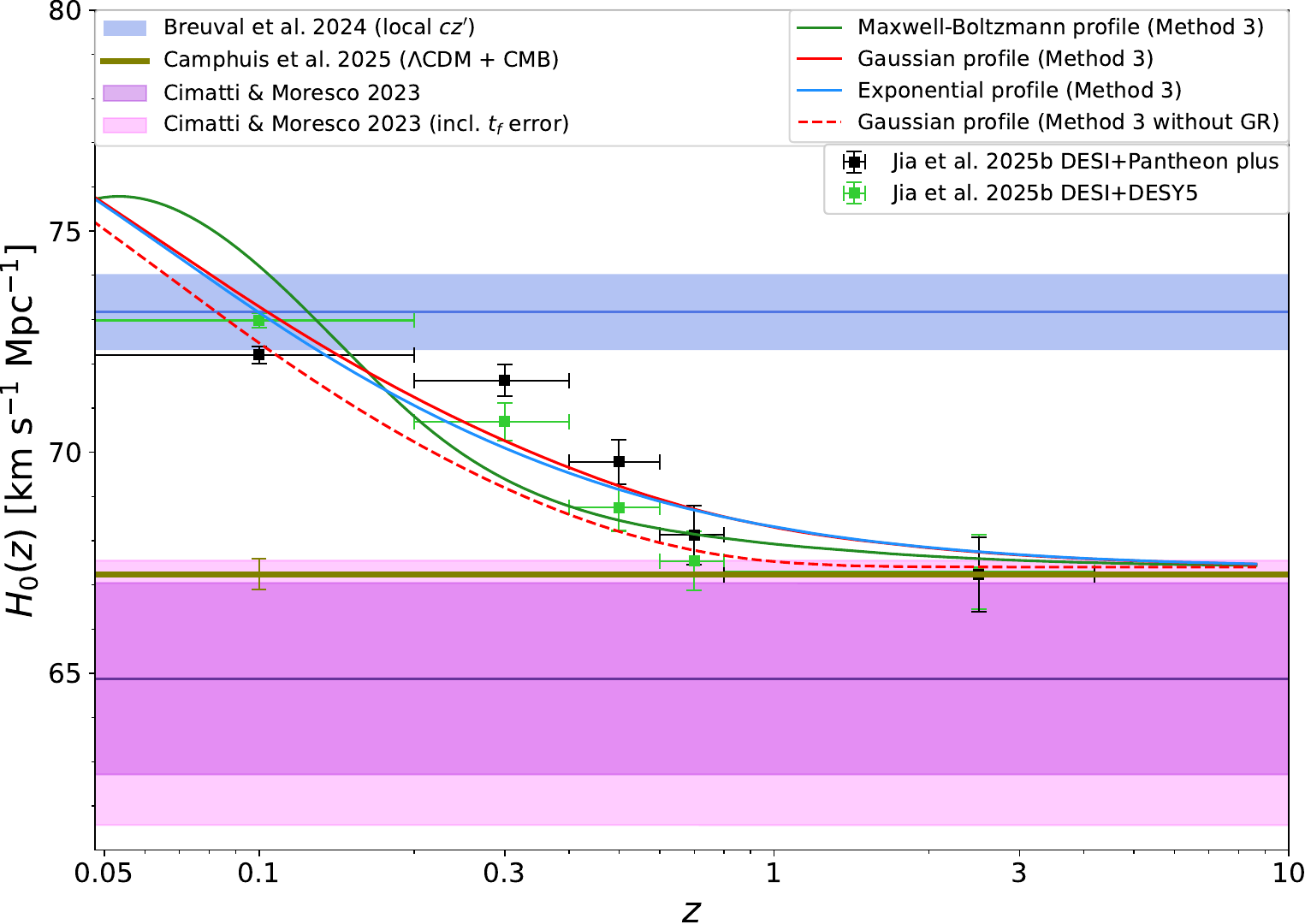}
    \caption{Predicted $H_0(z)$ curves in the void models and according to observations \citep{Jia_2025b}. The horizontal blue band at the top shows the local $cz'$ measurement \citep{Breuval_2024}, while the lower magenta band shows $H_0$ estimated from old Galactic stars and stellar populations \citep{Cimatti_2023}. The horizontal olive line shows $H_0^\mathrm{CMB}$, with uncertainty indicated at the left \citep{Camphuis_2025}. The solid curves show void model predictions \citep{Mazurenko_2025}. For illustrative purposes, the dashed red curve shows the prediction of the Gaussian model without including GR (Equation~\ref{z_contributions}). Its importance is clear from comparison with the solid red curve. The square points with uncertainties are based on a flexible reconstruction of the expansion history using BAO + uncalibrated SNe~Ia from Pantheon+ (black) or DES~Y5 (green). The horizontal uncertainties indicate the redshift range of each bin \citep{Jia_2025b}. Adapted from figure~3 of \citet{Mazurenko_2025}.}
    \label{H_z_graph}
\end{figure*}

The Hubble tension can be thought of as a mismatch between $H_0$ estimated from data at different redshifts. This can be quantified through the concept of $H_0(z)$, the value of $H_0$ that would be inferred by observers using data in a narrow redshift range centred on $z$ assuming the $\Lambda$CDM model \citep{Krishnan_2020, Krishnan_2021, Jia_2023, Jia_2025a, Jia_2025b}. Figure~\ref{H_z_graph} shows the predictions of the void models discussed above \citep{Mazurenko_2025}. Those authors considered three methods for how to extract $H_0(z)$ from $\dot{a}_{\mathrm{app}} (z)$. We adopt their Method~3, which they considered most comparable to how observers infer $H_0(z)$ (their other methods give similar results). The solid lines show predictions of the void models. The dashed red line shows the prediction of the Gaussian void profile without GR, highlighting the much faster return to $H_0^\mathrm{CMB}$ if this essential aspect of the model is neglected (Equation~\ref{z_contributions}). The horizontal blue band shows the local $cz'$ \citep{Breuval_2024}. The magenta band shows $H_0$ from cosmology-independent estimates of $A_\mathrm{U}$ using the ages of old Galactic stars \citep{Cimatti_2023}. The dark magenta band assumes $t_\mathrm{f} = 0.2$~Gyr, while the light magenta extension allows $t_\mathrm{f}$ to be at most $2\times$ smaller or larger. The solid olive line with error bar towards the left shows $H_0^\mathrm{CMB}$ \citep{Camphuis_2025}.

Figure~\ref{H_z_graph} compares the predicted $H_0(z)$ curves to recent observational estimates \citep{Jia_2025b}. These are shown as square points, with horizontal uncertainties indicating the redshift range of the BAO + uncalibrated SNe~Ia data they rely on. Correlations between redshift bins were carefully removed by diagonalising the covariance matrix. The black points use the Pantheon+ catalogue of SNe~Ia, while the green points use instead Dark Energy Survey Year 5 (DES~Y5). $\Lambda$CDM is assumed valid at early times, leading to a standard $r_\mathrm{d}$. Since there is no BAO anomaly at high redshift (Figure~\ref{D_V_graph}), the inferred $H_0(z) = H_0^\mathrm{CMB}$ at $z \ga 1$. $H_0(z)$ is reconstructed at lower $z$ using a flexible expansion rate history in which $w$ is allowed to have different values in different redshift bins, though $w$ remains constant inside each bin \citep{Jia_2025b}. The resulting expansion history need not match $\Lambda$CDM, but the authors still use this model to extrapolate the results to $z = 0$ and thereby obtain $H_0(z)$ (see their equation~19).

The use of uncalibrated SNe~Ia makes it remarkable that the predicted $H_0(z)$ agrees with the local $cz'$ as $z \to 0$ (Figure~\ref{H_z_graph}). This is presumably due to the growing BAO anomaly at low redshift (Figure~\ref{D_V_graph}) and a similar descending trend in $H_0(z)$ evident in SNe~Ia alone \citep[see table~2 of][]{Jia_2025b}. These trends allow their study to find an expansion history that deviates from $\Lambda$CDM only at low redshifts yet can simultaneously explain the CMB and the local $cz'$, informed by BAO and uncalibrated SNe~Ia from either Pantheon+ or DES~Y5. The resulting empirically determined $H_0(z)$ curve agrees very well with the prediction of the local void model, regardless of its assumed initial density profile.

The local $cz'$ plausibly matches the predicted $H_0$ assuming $\Lambda$CDM at high $z$, but with an empirically constrained expansion history at lower $z$ \citep{Jia_2025b}. In this approach, it is essential that the expansion history is not informed by the local $cz'$, as otherwise this quantity cannot be predicted. A complementary approach is to use the local $cz'$ as a constraint, but without the assumption of $\Lambda$CDM at high $z$ \citep{Lopez_2025}. Those authors used a wider range of datasets including CCs, but treated both BAO and SNe~Ia as uncalibrated. The local $cz'$ was fixed using purely geometric distances to megamaser host galaxies \citep{Pesce_2020}. This starts off the inferred $H_0(z)$ curve at the local $cz'$, but it quickly decreases to $H_0^\mathrm{CMB}$. Remarkably, there is no evidence for a Hubble tension outside the lowest redshift bin, with $H_0(z)$ at $z \ga 0.1$ much closer to $H_0^\mathrm{CMB}$ than to the local $cz'$ (see their figure~1). Treating both SNe~Ia and BAO as UCS makes their results less precise, but also far less model-dependent.

The consistent downward trend in $H_0(z)$ observed in both studies strengthens the case for a late-time deviation from $\Lambda$CDM, large enough that no new physics appears necessary at early times \citep{Jia_2025b, Lopez_2025}. A CMB-consistent expansion history at $z \ga 0.3$ is supported by measurements of $\gamma$-ray attenuation by extragalactic background light (EBL), which imply $H_0 = 62 \pm 4$~km/s/Mpc \citep{Dominguez_2019, Dominguez_2024}. We can also constrain $H_0$ using time delays between multiple images of strongly lensed SNe, which are far better suited to cosmology than quasars due to SNe being much more compact in spacetime \citep{Refsdal_1966}. Currently the most accurate such constraints come from SN Refsdal, which has been extensively studied over many years \citep{Kelly_2015, Kelly_2023, Grillo_2024, Liu_2025}. The results indicate that $H_0 = 66 \pm 4$~km/s/Mpc, in good agreement with $H_0^\mathrm{CMB}$ and the void model predictions at the lens galaxy redshift of 0.54 (Figure~\ref{H_z_graph}).

\section{Future tests}
\label{Future_tests}

\subsection{Galaxy number counts}
\label{Galaxy_number_counts}

Several studies point towards a local underdensity in source number counts across nearly the entire electromagnetic spectrum (Section~\ref{Local_void_solution}). Future observations and analyses can refine this picture further, checking assumptions such as a fixed shape to the galaxy luminosity function and using a more advanced statistical methodology, as well as constraining our position in the void through anisotropy in the galaxy number counts. The increased depth of more modern surveys should help to better fix the flat density region at $z \ga 0.1$, where the density appears to no longer increase with $z$. This region is presumably beyond the local void, but it is also somewhat challenging to observe \citep{Keenan_2013}. This could explain why a more recent study that only went out to $z = 0.08$ claimed a much shallower underdensity in redshift space of only $20 \pm 2\%$ \citep{Wong_2022}. Their own figure~1 shows that the comoving galaxy number density is still rising steeply by $z = 0.08$, indicating the study is unable to probe beyond the local void. This caused the authors to underestimate the cosmic mean density, and therefore the extent to which local galaxy number counts fall below it. The cosmic mean density appears to be about $1.5\times$ higher than they assumed, since it becomes flat just beyond $z = 0.08$ \citep{Keenan_2013}. Once this is accounted for, the local underdensity in redshift space is consistent between both studies \citep{Banik_2025_BAO}. A more careful quantification of the comoving galaxy luminosity density in the nearby Universe and its redshift dependence is currently underway (Pierce et al., in preparation).

\subsection{Peculiar velocities}
\label{Peculiar_velocities}

A local void would also create significant peculiar velocities and bulk flows in the nearby Universe \citep{Haslbauer_2020}. This appears to be in line with the anomalously fast bulk flows in CF4 \citep{Watkins_2023, Whitford_2023}. The predicted bulk flow curve in the void model depends on our location within the void as well as the void parameters \citep{Mazurenko_2024}. It is possible to test the void model at the field level by comparing it to the distances and redshifts of individual CF4 galaxies \citep{Stiskalek_2025_void}. Those authors found that when allowing the void size and depth to vary, the preferred parameters differed from the original ones \citep{Haslbauer_2020}. In particular, a smaller and potentially deeper void provided a better fit to the data, though results depend on the adopted density profile, growth history, and distance systematics. Importantly, even though the void models sit on a background \emph{Planck} cosmology, the predicted local $cz'$ can be as high as 73~km/s/Mpc at $1\sigma$ confidence with an exponential profile. While it is not presently clear whether the required void properties are in agreement with other datasets, this demonstrates that the $z < 0.05$ velocity field traced by CF4 does not obviously exclude the local void scenario for the Hubble tension. Future galaxy surveys should reach deeper while hopefully also improving their sky coverage and individual redshift-independent distances, which are still quite uncertain at present. This will help to refine the bulk flow curve and allow more detailed tests of the void model at the field level. It will also give us a better idea of our location within the void, since increasing offsets from its centre lead to greater anisotropy in the galaxy luminosity density and $cz'$. It would also typically increase the peculiar velocity of the Local Group in the CMB rest frame, which has been used to place constraints on our location within the void \citep{Haslbauer_2020, Mazurenko_2024, Stiskalek_2025_void}. While the observed 630~km/s \citep{Kogut_1993} is somewhat low compared to typical locations within the void, it is plausible at the 2\% level \citep{Haslbauer_2020}. For locations which satisfy this constraint, anisotropic lensing of the CMB by the KBC void remains within current limits \citep{Stiskalek_2025_void}.

\subsection{Fast radio bursts (FRBs)}
\label{FRBs}

Many of the successes of the local void scenario also carry over to background solutions to the Hubble tension that modify the Friedmann equations primarily at late times \citep{Najera_2026}. Both solutions have little impact on $A_\mathrm{U}$, so measurements of it cannot easily shed much light on this problem beyond excluding purely early time solutions. Distinguishing a local void from a background solution is difficult using an isotropically averaged relation between redshift and angular diameter or luminosity distance. Unless there is an extremely fundamental problem with cosmology, these types of distances can be related to each other through the Etherington relation \citep{Etherington_1933}. Little information is added by CC measurements of $\dot{z}$, since changes with respect to distance or lookback time contain equivalent information if $c$ has remained constant.

To test for the existence of a local void, one would ideally measure the mean matter density along the LOS to sources with spectroscopically confirmed redshifts. This is just the kind of information that can be provided by FRBs, enigmatic transient sources detectable in radio frequencies of roughly $0.1-8$~GHz \citep{Glowacki_2026}. These were first discovered in 2007 \citep{Lorimer_2007}. FRBs can often be localised to a very narrow region of the sky, linking them to a particular galaxy and thereby enabling a measurement of $z$ \citep{Pastor_2026}.

The novel aspect of an FRB lies in how its distance is constrained. FRBs are neither standard candles nor standard rulers. Their distance is inferred indirectly from the difference in arrival time of the low and high frequency components. As the FRB propagates from the source to Earth, it travels through ionized plasma containing free electrons. Their integrated effect along the entire Earth-source LOS delays the pulse arrival time. Importantly, the lower frequency components are delayed by a greater amount because their frequency is closer to the (density-dependent) plasma frequency, leading to a larger decrease in the FRB group velocity. As a result, the arrival time delay scales linearly with the integrated column density of electrons and inversely with the square of the observing frequency. This means measurements of the delay directly yield the column density of free electrons integrated over the source-detector line segment. This information is reported as the dispersion measure (DM).

The expansion of the Universe has been evident at optical wavelengths for $>100$~years from the fact that more distant galaxies have higher $z$ \citep{Wirtz_1922, Wirtz_1924}. The FRB analogue to this correlation is known as the Macquart relation \citep{Macquart_2020}. If we assume that $cz'$ is close to 70~km/s/Mpc, the slope of the Macquart relation has already been used to quantify the cosmic baryon density in the intergalactic medium \citep[IGM;][]{Connor_2025}. When combined with the baryon content in bound structures like galaxies and galaxy clusters, the total is in very good agreement with expectations from the CMB and Big Bang nucleosynthesis \citep{Cyburt_2016}. This paves the way for FRBs to be used as precision cosmological tools.

The Macquart relation is particularly useful in the context of a local void. If DM is converted into a distance assuming homogeneity but we are actually living in a supervoid, the distance will be underestimated. This will lead to an overestimated $cz'$. Therefore, FRBs should give a higher estimate of $H_0$ than other techniques based on standard candles, rulers, or clocks, all of which would remain accurate even if we are in a supervoid. Recent FRB estimates of $H_0$ are indeed somewhat higher than the typically measured local value, with one study reporting around $76-77$~km/s/Mpc using various empirically motivated assumptions about DM\textsubscript{host}, the DM contributed by the FRB host galaxy \citep{Kalita_2025}. Although their reported uncertainties are almost 3~km/s/Mpc, this will only get better given the rapidly increasing number of localised FRBs \citep{Kalita_2026}. Some of them also have a measurement of scattering time, which better constrains DM\textsubscript{host}. Including this information further tightens the constraints \citep{Yang_2025}.

To get a rough estimate of how much the FRB-derived $H_0$ might exceed other estimates, we assume a median FRB redshift of 0.3 \citep{Kalita_2025}. A 20\% underdensity out to $z = 0.1$ would imply a roughly $1/30$ reduction in the DM.\footnote{Comparing the integral of unity from $z = 0-0.3$ with the integral of a model where the integrand rises linearly from 0.8 to 1 over $z = 0-0.1$ and remains flat thereafter.} This would approximately increase $H_0$ by the same amount. If the local $cz'$ is taken to be 73.5~km/s/Mpc, we might expect FRBs to give instead 76~km/s/Mpc. This is remarkably close to the actual result reported by \citet{Kalita_2025}.

The somewhat higher $H_0$ estimated from FRBs compared to more traditional observables motivates a more detailed analysis accounting for the present density profile of a local void and the distortions it creates to the distance-redshift relation. This is because in addition to the reduced DM arising from a local underdensity, we would still have extra non-cosmological contributions to the redshift of the FRB host galaxy (Equation~\ref{z_contributions}). Only the latter would affect more conventional probes of the distance-redshift relation. By comparing the results, it might even be possible to constrain both the depth of the void and its present peculiar velocity field, which is related to how quickly it is getting deeper. This in turn would tell us something about the behaviour of gravity on $\ga 100$~Mpc scales. Taking the KBC void at face value, some modification to the standard growth history seems likely \citep{Haslbauer_2020}. A careful combination of cosmological probes including FRBs may provide an independent test of this.

\subsection{The redshift drift experiment}
\label{Redshift_drift_experiment}

The datasets considered so far are all instantaneous measurements along our past lightcone, which we can normally think of as infinitely thin. Recording data over an extended period of time allows us to take cosmological measurements over a thick past lightcone, giving access to qualitatively new information. This is the basic idea behind the redshift drift experiment \citep{Sandage_1962, McVittie_1962, Loeb_1998}. The redshift of any object will change over time because after waiting some period $\Delta t_\mathrm{obs}$ at Earth, we would see photons which were emitted from the source some time $\Delta t_\mathrm{emit} = \Delta t_\mathrm{obs}/(1 + z)$ later, essentially from the very definition of redshift \citep{Trost_2025}. Accounting for cosmic expansion at both the present Earth and the time of emission from the source, those authors obtained the classic result that
\begin{eqnarray}
    \dot{z} ~=~ H_0 \left( 1 + z \right) - H \left( z \right) \, ,
    \label{Redshift_drift_rate}
\end{eqnarray}
where the overdot denotes a time derivative with respect to when the measurement is taken.

Equation~\ref{Redshift_drift_rate} assumes homogeneous expansion. In the local void solution, $\dot{z}$ could in principle be obtained by differentiating Equation~\ref{z_contributions}, but this would entail significant complications. We instead differentiate Equation~\ref{z_contributions} numerically with respect to the present epoch. For this, we calculate the redshift of each test particle when its spacetime trajectory intersects our past lightcone. We then repeat this calculation with the cosmic age at the present epoch shifted by $\pm 1$~Myr, since the redshift drift $\Delta z$ over a few years is not well resolved numerically.\footnote{1~Myr is still much shorter than the light crossing time of a local supervoid, so we expect this procedure to be accurate.} Observers at a future epoch would perceive a slightly lower $H_0$ and $\Omega_\mathrm{M}$. For the trajectory integrations to start at the same time relative to the Big Bang, the starting redshift must be increased slightly. The same applies to the initial comoving radius, since we wish to keep the same initial physical distance from the void centre in order to track the same particle. Given its apparent redshift at different observing times, we use a centred differencing approach to compute $\dot{z}$, which we then use to calculate $\Delta z$ over a more realistic observing period of a decade.

\begin{figure*}
    \centering
    \includegraphics[width=0.49\textwidth]{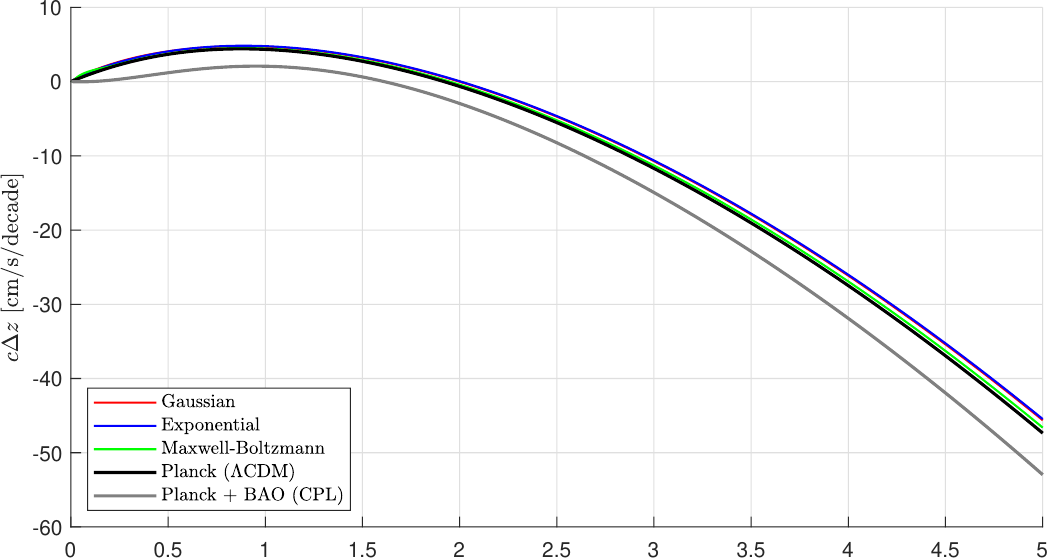}
    \hfill
    \includegraphics[width=0.49\textwidth]{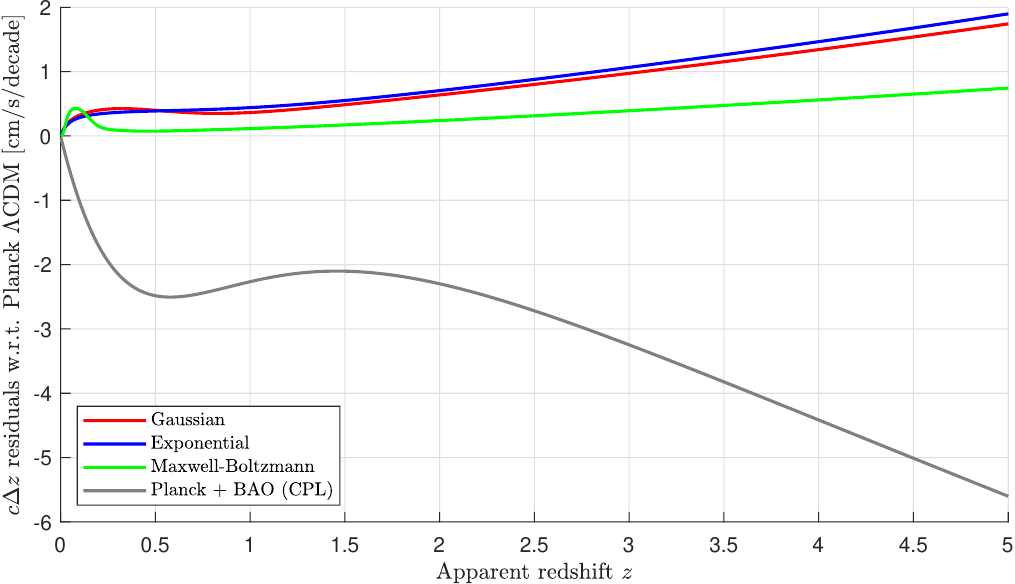}
    \caption{The forecasted redshift drift $\Delta z$ over a decade in different cosmological models. The solid black line in the left panel shows the homogeneous \emph{Planck} cosmology, which is subtracted from the other models in the right panel to highlight differences. The coloured curves slightly above it show predictions in the void models \citep{Haslbauer_2020}. The low grey curve shows the CPL model fit to \emph{Planck} CMB + BAO \citep{Mirpoorian_2025}.}
    \label{zdrift_graph}
\end{figure*}

The model predictions are shown in Figure~\ref{zdrift_graph}. The left panel shows the forecasted change in redshift over a decade, with the void models shown using coloured lines. The solid black line shows the homogeneous \emph{Planck} cosmology, while the grey line shows the CPL model calibrated using CMB + BAO \citep{Mirpoorian_2025}. The right panel better highlights differences between these models by subtracting the prediction of the homogeneous \emph{Planck} cosmology. The largest residuals arise in the CPL model, but even this model has an expected signal of only a few cm/s. There are almost certainly far easier ways to distinguish the homogeneous CPL and \emph{Planck} cosmologies. However, in the local void solution, the redshift drift experiment offers a chance to observe in real time how the peculiar velocity and gravitational redshift change, not merely how their present values distort the distance-redshift relation along our past lightcone.

The right panel of Figure~\ref{zdrift_graph} shows that our considered void models can only be distinguished from the homogeneous \emph{Planck} cosmology with a velocity precision of a few mm/s over a decade. Clearly, the redshift drift experiment is still far from yielding cosmologically meaningful results. Forecasts indicate that it may be feasible using multi-decade observations with instruments currently under construction \citep{Alves_2019, Esteves_2021, Rocha_2023, Trost_2025}. This would offer an interesting new probe of cosmology, allowing tests of homogeneity by comparing $\dot{z}$ measurements at the same $z$.

\subsection{The kinematic Sunyaev-Zel'dovich (kSZ) effect}
\label{kSZ_effect}

The local void solution differs from background solutions to the Hubble tension in whether this tension exists for all observers at the present epoch. If it does not, significant peculiar velocities are essential. Peculiar velocities of individual objects are difficult to access directly because it is not usually possible to disentangle the various contributions to the total redshift (Equation~\ref{z_contributions}). Fortunately, this is possible to a limited extent using the kSZ effect \citep{Sunyaev_1980}. Electrons in hot gas in galaxy clusters can scatter CMB photons, changing their energy and therefore the observed CMB temperature towards the cluster. This allows us to measure its LOS velocity relative to the CMB. The kSZ effect might therefore permit a test of the large outward peculiar velocities predicted by the local void solution \citep{Mazurenko_2024, Stiskalek_2025_void}. The outflows that might arise from a KBC-like void are well within current observational limits \citep{Ding_2020, Cai_2025}.

\subsection{Structure growth on very large scales}
\label{LSS_growth}

The formation of the KBC void and its proposed impact on the local $cz'$ require faster growth of structure than is possible in $\Lambda$CDM, which predicts a cosmic variance in $cz'$ measurements over the commonly used $100-600$~Mpc distance range of only about 0.9~km/s/Mpc \citep{Wu_2017, Camarena_2018}. If structure growth on $\ga 100$~Mpc scales is indeed faster than in $\Lambda$CDM, it should manifest in other ways. It is currently very difficult to test this because even the density profile of the KBC void remains difficult to constrain \citep{Wong_2022}. At larger distances, only the brightest galaxies would remain detectable. These might be biased tracers of the total matter distribution, creating a degeneracy between how much the density fluctuates and the level of bias of the tracer used \citep{Haslbauer_2020}.

There are already a few hints that structure growth might indeed be faster than in $\Lambda$CDM on very large scales. One example is the El Gordo galaxy cluster collision, whose extremely high mass, collision velocity, and redshift of 0.87 is in $>5\sigma$ tension with $\Lambda$CDM for any plausible infall velocity that reproduces its observed morphology in existing hydrodynamical simulations \citep*{Asencio_2021, Asencio_2023}. A supervoid similar to the KBC void might also be responsible for the CMB Cold Spot \citep{Nadathur_2014, Kovacs_2020, Kovacs_2022_Cold_Spot}. The CMB power spectrum on large scales (low multipole moments) is sensitive to the late-time integrated Sachs-Wolfe (ISW) effect \citep{Sachs_1967}. This allows the CMB power spectrum to break a degeneracy between $H_0$ and the CMB monopole temperature $T_0$ \citep{Yoo_2019, Ivanov_2020_TCMB}. If $\Lambda$CDM worked correctly, we might expect such a procedure to recover the standard $H_0^\mathrm{CMB}$ and be consistent with the measured value of $T_0 = 2.72548 \pm 0.00057$~K from the Far-Infrared Absolute Spectrophotometer (FIRAS) instrument onboard the Cosmic Background Explorer (COBE) satellite \citep{Fixsen_2009}. However, analysis of \emph{Planck} data without the usual FIRAS prior on $T_0$ implies that $T_0 = 3.144^{+0.17}_{-0.065}$~K and $H_0 = 55.32^{+1.7}_{-4.9}$~km/s/Mpc, with little room for a dark energy component \citep*{Ivanov_2020_TCMB}. If the assumption of a flat Universe is relaxed, the data actually prefer a closed Universe at high confidence \citep{Valentino_2020_flat}. While \citet{Ivanov_2020_TCMB} pass off the unusually low $H_0$ and high $T_0$ as a statistical fluctuation, their analysis already includes uncertainties in the CMB power spectrum on large angular scales and cosmic variance. Since their figure~3 shows that moving along the $H_0 - T_0$ degeneracy direction only affects large angular scales once other parameters are appropriately adjusted, something about the data here drives the preference for a low $H_0$. The most plausible culprit is the anomalously low power on these scales \citep{Das_2014}, which can to some extent be fit with a high $T_0$ and low $H_0$ \citep[figure~3 of][]{Ivanov_2020_TCMB}. We suggest that this is indicative of a problem for $\Lambda$CDM with the growth of LSS and the resulting ISW effect. Indeed, it has been argued that the ISW signal from stacked supervoids exceeds the $\Lambda$CDM expectation \citep{Kovacs_2022_ISW}.

Future observations should eventually clarify if cosmic variance in $cz'$ on the several hundred Mpc scale used to measure its local value might be sufficient to solve the Hubble tension, a possibility that is not compatible with $\Lambda$CDM. Given the difficulties in characterising the KBC void, it may be some time before observations cover a sufficiently large volume in sufficient depth to enable a statistical analysis of KBC-like voids, even if they are common. Alternatively, such voids may remain intrinsically rare, even in cosmologies with faster-than-$\Lambda$CDM structure growth \citep{Samaras_2025}.

The growth of structure on large scales might also be constrained using the KBC void itself, once its density and velocity field are better understood \citep{Malandrino_2026, Stiskalek_2026_GA}. Combining FRBs with other cosmological probes may give some insight into the present rate at which the void is getting deeper. This is because FRBs probe the combined effect of the local underdensity and the void-induced distortion to the distance-redshift relation -- but other probes are usually sensitive only to the latter (Section~\ref{FRBs}). This could allow a determination of both the density contrast and the outflow velocity, perhaps providing a novel window into the behaviour of gravity on otherwise poorly probed $\ga 100$~Mpc scales.

\section{Summary and conclusions}
\label{Conclusions}

There is now growing evidence that redshifts in the local Universe increase with distance about 9\% faster than expected in the $\Lambda$CDM paradigm calibrated using the CMB \citep{H0DN_2025, Valentino_2025}. This statistically significant Hubble tension indicates significant deficiencies in our understanding of cosmology. If the solution lies in higher present $\dot{a}$ at the background level (preserving Equation~\ref{Hubble_law}), our understanding of dark energy will likely need to be modified, which may have profound implications for the ultimate fate of the Universe. Background solutions can also arise through modifications to $\Lambda$CDM prior to recombination, which in principle could allow an adequate fit to the CMB at higher $H_0$. However, early time solutions are unlikely for several reasons \citep{Lin_2021_UCS, Vagnozzi_2023, Banik_2025_cosmology}. In addition to difficulties fitting the CMB power spectrum, these scenarios generically predict that the Hubble tension persists out to high redshift, implying that probes at $z \la 1100$ can all be understood in $\Lambda$CDM with the same values of the cosmological parameters. Figure~\ref{OmegaM_h} shows that this is not the case, since in reality there is agreement with \textit{Planck} $\Lambda$CDM in cosmological probes both before and after recombination -- except for the outlying local $cz'$.

Another strong argument for a low redshift solution to the Hubble tension is the nearly $3\sigma$ BAO anomaly recently reported in DESI~DR2 \citep{DESI_2025}. There is no BAO anomaly at $z \ga 1.5$ despite the excellent precision (Figure~\ref{D_V_graph}). However, a growing anomaly appears at $z \la 1$, an issue that cannot be fully addressed solely by changing $r_\mathrm{d}$ through new physics prior to recombination. The BAO anomaly is ultimately a mismatch between BAO observables and their predicted values in $\Lambda$CDM calibrated using the CMB. If one also uses the local $cz'$ as an extra constraint on the $\Lambda$CDM parameters, then the BAO data can be predicted remarkably well \citep{Najera_2026}. Although $\Lambda$CDM cannot jointly fit the CMB and the local $cz'$, the correct cosmological model will achieve a compromise solution that adequately fits both. It therefore seems plausible that the correct expansion history is approximately given by fitting $\Lambda$CDM to both the CMB and the local $cz'$. The ability of $\Lambda$CDM with parameters constrained in this way to predict the BAO data makes it unlikely that the local $cz'$ is measured incorrectly -- if it was, its high precision would skew the $\Lambda$CDM parameters to incorrect values, most likely leading to incorrect BAO predictions. Therefore, the Hubble and BAO tensions faced by $\Lambda$CDM reinforce each other, painting a consistent picture in which distances to a given redshift are lower than expected only at low $z$, but become consistent with $\Lambda$CDM at higher $z$.

Long before these recent tensions became known, a local underdensity or void was proposed based on optical galaxy number counts \citep{Maddox_1990, Shanks_1990}. Given the nearly uniform initial conditions evident in the CMB, a local void can only have arisen through outflows beyond cosmic expansion alone. Since the void would still be developing today, the continuity equation implies net outward peculiar velocities \citep{Haslbauer_2020}. Those authors made a simple estimate that these would inflate $cz'$ by about $11 \pm 2\%$ given the apparent underdensity of the local void in redshift space galaxy counts (Equation~\ref{H0_estimate_void}). By setting up a semi-analytic model of a slightly underdense spherical void at high redshift and integrating it forward to $z = 0$, those authors related the void parameters to key cosmological observables -- but they did not consider BAO data. Their best-fitting model parameters can therefore be used to predict the BAO data \citep{Banik_2025_BAO}. The results reveal good agreement with available BAO measurements over the last 20 years (Figure~\ref{D_V_graph}). Indeed, the fits are far better than the homogeneous \emph{Planck} cosmology, the \emph{a priori} prediction of $\Lambda$CDM at the time.

The local void solution predicts that the Hubble tension decays at high redshift in a specific way \citep{Mazurenko_2025}. This can be quantified by $H_0(z)$, the $H_0$ that would be inferred from data in a narrow redshift range centred on $z$ if the measurements are extrapolated to $z = 0$ assuming a $\Lambda$CDM expansion history \citep{Jia_2023, Jia_2025a}. The predicted $H_0(z)$ curve appears to agree very well with the latest determinations (Figure~\ref{H_z_graph}), though it is important to quantitatively compare the void model to the reconstructed distance--redshift relation. A remarkable aspect of the observational results is that assuming a flexible expansion history constrained by BAO and uncalibrated SNe~Ia, it is possible to reconcile the assumption of $\Lambda$CDM at high redshift with the locally measured $cz'$ \citep{Jia_2025b}. Similar results can be obtained treating both BAO and SNe~Ia as uncalibrated but adding a purely geometric local measurement of $cz'$ as an extra constraint \citep{Lopez_2025}. Their figure~1 illustrates that beyond the lowest redshift bin, the Hubble tension is not present and $H_0(z)$ is far more in line with $H_0^\mathrm{CMB}$ than it is with the local $cz'$. Further improvements are possible today by updating the BAO data from DESI~DR1 to DESI~DR2 \citep{DESI_2025}.

Many of these successes are not unique to the local void solution -- a purely background solution to the Hubble tension would also predict a BAO anomaly only at low $z$ and some form of descending trend in $H_0(z)$ \citep{Najera_2026}. In Section~\ref{Future_tests}, we discuss how it may be possible to distinguish between a local void and a background solution at late times. The most promising technique in the near term seems to be the Macquart relation of FRBs, whose DM is sensitive to the integrated number density of free electrons along the LOS to the source (Section~\ref{FRBs}). The general expectation is that if homogeneity is assumed when converting DM to distance, the distance will be underestimated from within a local void, leading to an overestimated $H_0$. This would create a few percent difference between $H_0$ estimated from FRBs and from other sources based on standard candles, rulers, or clocks, all of which would continue to give accurate distances (or lookback times) even if we are in a supervoid. Recent FRB results do indeed suggest a higher $H_0$ than the great majority of estimates for the local $cz'$ \citep{Kalita_2025}. Further investigation of this issue may be able to confirm the existence of a local void or rule it out, taking advantage of the rapidly increasing number of localised FRB detections \citep{Kalita_2026, Pastor_2026}. Combining FRBs with other cosmological probes may even give some insight into the present rate at which the void is getting deeper. Other tests of a local void are also possible, for instance using the kSZ effect (Section~\ref{kSZ_effect}) or the futuristic redshift drift experiment (Section~\ref{Redshift_drift_experiment}). One should also not overlook incremental advances to existing techniques, especially continued investigation of galaxy number counts and peculiar velocities at $z \la 0.2$ (Sections~\ref{Galaxy_number_counts} and \ref{Peculiar_velocities}, respectively).

The local void solution is relatively unique among solutions to the Hubble tension in that it was not originally proposed for this purpose. Constraining its parameters by requiring it to do so leads to a plethora of predictions for observations not used to fix the void parameters \citep{Haslbauer_2020}. These predictions have largely been successful so far, especially for BAO measurements over the last 20~years (Figure~\ref{D_V_graph}). The model naturally explains why solutions relying solely on new physics prior to recombination now appear unlikely (Figure~\ref{OmegaM_h}). While many of these successes could in principle be attributed to a late-time background solution to the Hubble tension \citep{Najera_2026}, such models do not naturally account for the presence of the KBC void. By contrast, the local void solution self-consistently links the KBC void and the Hubble tension as cause and effect. Confirming the void hypothesis requires distinguishing it from a late-time background solution, a task that remains challenging but may be addressed using some of the techniques outlined in Section~\ref{Future_tests}. This is crucial to unlocking the rich theoretical insights that seem increasingly likely to emerge from the increasingly significant Hubble tension \citep{H0DN_2025, Valentino_2025}. Whatever its solution ultimately turns out to be, science progresses by taking anomalies seriously, the Hubble and BAO tensions not least among them.

\section*{Acknowledgements}

IB and HD are supported by Royal Society University Research Fellowship grant 211046.

%Thank referee.

\section*{Data Availability}

No new data were created or analysed for this contribution.

\bibliographystyle{mnras}
\bibliography{NVR_bbl}
\bsp %Typesetting comment. Do not change this block of three lines.
\label{lastpage}
\end{document}